\documentclass[amsmath,11pt]{article}
\usepackage{amsfonts,amsmath,amssymb,graphics,epsfig}
\usepackage{enumerate}\usepackage{theorem}
\usepackage{xcolor}

\date{\empty}

\title{Could electromagnetism be envisaged as a form of gravity in a metric affine framework?}

\author{Panagiotis Mavrogiannis\\ {\small Section of Astrophysics, Astronomy and Mechanics, Department of Physics}\\ {\small Aristotle University of Thessaloniki, Thessaloniki 54124, Greece}}

\begin{document}

\maketitle

\begin{abstract}
We revisit the relativistic coupling between gravity and electromagnetism, putting particularly into question the status of the latter; whether it behaves as a source or as a form of gravity on large scales. Considering a metric-affine framework and a simple action principle, we find out that a component of gravity, the so-called homothetic curvature field, satisfies both sets of Maxwell equations. Therefore, we arrive at a gravito-electromagnetic equivalence analogous to the mass-energy equivalence. We raise and discuss some crucial questions implied by the aforementioned finding, refreshing our viewpoint of electromagnetism in curved spacetime.
\end{abstract}

\section*{Introduction}\label{sec:Introduction}

There are two kinds of well-known (fundamental) macroscopic field quantities introduced to causally describe the motion of matter on large (macroscopic) scales. These are the gravitational and electromagnetic fields, which are conventionally described by General Relativity and Maxwellian Electromagnetism. Due to the wide presence of electromagnetic fields in astrophysical and cosmological environments, we frequently need to consider the parallel presence, coupling or coexistence of gravity and electromagnetism on large-scales. In practice, our conventional perspective consists of envisaging electromagnetic fields (in analogy with matter fields) as sources of gravitation, and therefore generalising the laws of electrodynamics to curved spacetimes (we talk about electrodynamics in curved spacetime). However, unlike (ordinary) matter fields\footnote{There are in fact two explicitly known forms of matter (taking into account the mass-energy equivalence), `ordinary' matter and electromagnetic fields. These can be described by scalar and vector/tensor fields respectively.}, electromagnetic ones possess a geometric nature which allows for their double coupling with spacetime curvature, not only (indirectly) via Einstein's equations but (directly) through the so-called Ricci identities as well\footnote{Apart from the so-called \textit{Einstein-Maxwell coupling}, the \textit{Weyl-Maxwell coupling} (long-range curvature and electromagnetic field) has also been studied within the literature~\cite{MB}.}~\cite{T11}. Let us have a closer look at the two aforementioned types of coupling.

\subsection{Einstein-Maxwell coupling}\label{ssec:Einstein-Maxwell}

Firstly, according to Einstein's equations of gravitation, electromagnetic fields, as a form of energy, along with (ordinary) matter contribute to the formation of spacetime geometry. Hence, in a sense, Maxwell's electromagnetism is incorporated into (or makes part of) General Relativity by providing a kind of source for the gravitational field. Let us write here for reference the relativistic equations for gravity\footnote{We adopt the geometrised units system (refer to the appendix F in~\cite{W1} for details), i.e. $8\pi G=1=c$ ($G$ is the gravitational constant and $c$ is the speed of light), in which all quantities (ordinarily measured in terms of the fundamental units of length $L$, time $T$ and mass $M$) have dimensions expressed as integer powers of length. In particular, we note that mass, time, electric charge and energy have dimensions of length; velocity, force, action and Maxwell potential are dimensionless; Faraday electromagnetic field has inverse length dimensions whilst energy density and electric current density are measured in inverse square length units.},
\begin{equation}
    R_{ab}-\frac{1}{2}Rg_{ab}=T_{ab}\,,
    \label{eq-Einstein}
\end{equation}
where the (symmetric) Ricci tensor $R_{ab}$ encodes the local-gravitational field (the Ricci scalar $R=R^{a}{}_{a}$ provides a measure of the average local gravity), and the energy-momentum tensor $T_{ab}$ (Noether conserved currents under translations and rotations) represents the energy sources of gravitation. For ordinary matter and electromagnetic fields, the aforementioned tensor reads: $T_{ab}=T^{\text{(m)}}{}_{ab}+T^{\text{(EM)}}{}_{ab}=T^{\text{(m)}}{}_{ab}+F_{ac}F^{c}{}_{b}+(1/4)F_{cd}F^{cd}g_{ab}$, where $F_{ab}=2\partial_{[a}A_{b]}=\partial_{a}A_{b}-\partial_{b}A_{a}$ is the (antisymmetric) Faraday tensor, written in terms of the Maxwell 4-potential $A_{a}$. Finally, the metric (also symmetric) tensor $g_{ab}$ encodes the geometric properties of spacetime and is used to calculate lengths and angles in Riemannian manifolds. Through Einstein's equations $R_{ab}$ and $g_{ab}$ represent different aspects of the same beingness. In particular, $g_{ab}$ is an agent for forming the Ricci scalar, basically the Lagrangian of relativistic gravitation, and it does not refer to inherent properties of spacetime but it is determined by the coordinate description of the physical system in question

\subsection{Ricci-Maxwell coupling}\label{ssec:Ricci-Maxwell}

Secondly, due to their geometric (vector/tensor) nature, electromagnetic fields directly couple with spacetime curvature via the so-called Ricci identities. The aforementioned coupling is a purely geometrical interaction that goes beyond the usual interplay between matter and geometry monitored by Einstein's equations~\cite{T11}. In particular, the latter when applied to the Maxwell 4-potential and the Faraday tensor fields, read (in Riemannian geometry):
\begin{equation}
    2\nabla_{[a}\nabla_{b]}A_{c}=R_{abcd}A^{d} \hspace{15mm} \text{and} \hspace{15mm} 2\nabla_{[a}\nabla_{b]}F_{cd}=-2R_{ab[c}{}^{e}F_{d]e}\,,
    \label{eq-Ricci}
\end{equation}
respectively, where $R_{abcd}$ is the Riemann curvature tensor encoding the total spacetime curvature. The above relations differ from the gravitational field equations in that firstly, they permit the coupling of only geometric quantities with the spacetime curvature. Secondly, they associate geometric fields with the total spacetime curvature (involving the Weyl, long-range curvature, field as well) apart from the local one (as encoded by the Ricci tensor).

Overall, it seems to us that the aforementioned special coupling, described through~\eqref{eq-Ricci}, makes the status of electromagnetic fields essentially differ from that of a source of gravitation (e.g. ordinary matter). The aforementioned observation along with another, consisting of the mathematical similarity between the Faraday tensor $F_{ab}=2\partial_{[a}A_{b]}$ and the so-called \textit{homothetic curvature} tensor field $\hat{R}_{ab}=\partial_{[a}Q_{b]}$ (follow the subsequent discussion), motivated us to investigate whether electromagnetic fields could be envisaged as a form/component of spacetime curvature. 

\subsection{Metric affine framework}\label{ssec:Metric-affine}

Let us briefly present the metric-affine framework~\cite{I}, within which, the above mentioned field $\hat{R}_{ab}$ exists. To begin with, the transition from relativistic to metric affine spacetime requires rising two constraints of Riemannian geometry. On the one hand, we allow for an antisymmetric connection part, $S_{ab}{}^{c}\equiv\Gamma^{c}{}_{[ab]}$ (i.e. the torsion tensor); on the other hand, for a non-vanishing covariant derivative of the metric tensor, $Q_{abc}\equiv-\nabla_{a}g_{bc}\neq 0$ (note that $Q_{a}=g^{bc}Q_{abc}=Q_{ac}{}^{c}$ and $q_{a}=g^{bc}Q_{cba}=Q^{c}{}_{ca}$ are the non-metricity vectors). The former is associated with the impossibility to form infinitesimal parallelograms via parallel transport of a vector upon the direction of another and vice versa; the latter implies the vector length change during parallel transport. Within a metric-affine geometry, the Ricci tensor has also an antisymmetric part, containing contributions from both torsion and non-metricity. Homothetic curvature $\hat{R}_{ab}=\partial_{[a}Q_{b]}$ is just a component of that antisymmetric part. In the particular case of torsionless spacetime, one has $R_{[ab]}=\hat{R}_{ab}$. While Riemann curvature (or \textit{direction curvature}) is responsible for changes in the direction of parallelly transported vectors along a closed curve, homothetic curvature (or \textit{length curvature}) is associated with changes in vectors' length. It is worth noting that within the literature, the spacetime property of vectors' length change has been argued that it leads to the so-called \textit{second clock effect}, the exclusion of existence of sharp spectral lines, and therefore to a non-physical theory. In particular, the aforementioned problem dates back to Weyl's gauge theory of gravity and Einstein's associated objections (for some historical information refer to e.g.~\cite{HG}; for a modern approach to Weyl's theory see e.g.~\cite{SLFDR}). Interestingly however, it has been recently shown~\cite{HL1},~\cite{HL2} that under appropriate redefinition of proper time and the covariant derivative, the second clock effect does not actually arise in gravity theories with non-metricity.

Up to this point our aim may have already become clear. We will examine whether $\hat{R}_{ab}$ satisfies Maxwell equations, and whether there is a correspondence between homothetic curvature and the Maxwell field. In particular, it is the goal of the present manuscript to bring to the attention of the scientific community, the observation that there is indeed a (metric-affine) curvature component field which actually turns out to present an equivalence with the Maxwell field. In face of this finding we put into question our conventional perspective regarding the way we envisage macroscopic electromagnetic fields and their relation to gravity.\\

\section{Homogeneous (metric affine) Maxwell equations and the implication for gravito-electromagnetic equivalence}\label{sec:Homog-Maxwell-eqs}

Let us start from the expression $\nabla_{[a}F_{bc]}=(1/3)(\nabla_{a}F_{bc}+\nabla_{c}F_{ab}+\nabla_{b}F_{ca})$, within a Riemannian framework. According to the homogeneous Maxwell equations, it has to be equal to zero. Taking thus into account that the Faraday tensor comes from a potential 4-vector, we follow the operations:
\begin{eqnarray}
    \nabla_{[a}F_{bc]}&=&\frac{1}{3!}\left[2\nabla_{[a}\nabla_{b]}A_{c}+2\nabla_{[c}\nabla_{a]}A_{b}+2\nabla_{[b}\nabla_{c]}A_{a}\right]=\frac{1}{3}\left(R_{abcd}+R_{cabd}+R_{bcad}\right)A^{d} \nonumber\\&&
    =R_{[abc]d}A^{d}=0\,.
    \label{eq:Maxwell-Ricci}
\end{eqnarray} 
In other words, we have recalled that if a second-rank antisymmetric tensor field can be written as the gradient of a 4-vector field, then the homogeneous Maxwell equations are a consequence of two geometric properties of the Riemannian spacetime\footnote{Besides, the homogeneous Maxwell equations can be derived theoretically in Minkowski spacetime~\cite{LMEN1} through variation of the action $\mathcal{S}=\int{\left(-\sum{m_{i}}\sqrt{\eta_{ab}\dot{x}_{(i)}^{a}\dot{x}_{(i)}^{b}}-\frac{1}{4}F_{cd}F^{cd}-\sum{e_{i}}A_{a}\dot{x}_{(i)}^{a}\right)\,d\tau}$ with respect to the particles' coordinates $x^{a}(\tau)$ ($\tau$ is the particle's proper--time, its world--line parameter). Subsequently, the homogeneous Maxwell equations are generalised to curved (Riemannian) spacetime via the so--called minimal substitution rule.}; these are the Ricci identities in the form of~\eqref{eq-Ricci} and the first Bianchi identities (i.e. $R_{[abc]d}=0$).\footnote{For an arbitrary vector field $A_{a}$ the aforementioned properties imply that $\nabla_{[a}\nabla_{b}A_{c]}=0$.} Inversely, if the homogeneous Maxwell equations are satisfied, the second-rank antisymmetric tensor field can be written as the gradient of a 4-vector field in Riemannian spacetime. Therefore, it is clear that $\nabla_{[a}\hat{R}_{bc]}=0$ within the geometry in question. It is worth noting that the above well-known conclusion can be generalised to (non-Riemannian) geometries which possess non-metricity\footnote{It can be shown that both the Ricci and the first Bianchi identities maintain their Riemannian form when the relativistic background is modified by the additional non-metricity requirement. In fact, non-metricity is incorporated into the Riemann tensor.} (e.g. see eqs. (1.152) and (1.158) in~\cite{I}, corresponding to the metric-affine version of the Ricci identities and of the first Bianchi identities respectively). Nevertheless, in a general metric affine geometry, possessing torsion as well, the homogeneous Maxwell equations cease to be valid (once again see eqs. (1.152) and (1.158) of~\cite{I}, in combination with~\eqref{eq:Maxwell-Ricci}). In this case, homothetic curvature satisfies the following generalised version of Bianchi identities (known as \textit{Weitzenbock identities}-see eq (1.169) in~\cite{I})
\begin{equation}
    \nabla_{[a}\hat{R}_{bc]}=2\hat{R}_{d[a}S_{bc]}{}^{d}\,.
    \label{eqn:Weitzenbock}
\end{equation}
Observe that in the absence of torsion, $\hat{R}_{ab}$ satisfies the homogeneous set of Maxwell equations (i.e. $\nabla_{[a}\hat{R}_{bc]}=0$). Besides, it is known that Einstein-Hilbert action\footnote{In fact, there is a generalised action (known under the name \textit{quadratic theory}~\cite{I2}), containing the Einstein-Hilbert, which has as a consequence the property $S_{ab}{}^{c}=-(2/3)S_{[b}\delta_{a]}{}^{c}$.} implies that $S_{ab}{}^{c}=-(2/3)S_{[b}\delta_{a]}{}^{c}$ (with $S_{a}\equiv S_{ab}{}^{b}$ being one of the torsion vectors)-see~\cite{I}. Given the aforementioned property, let us point out that homothetic curvature satisfies (recall eq.~\eqref{eqn:Weitzenbock}) the following homogeneous set of Maxwell-like equations, namely 
\begin{equation}
    \textcolor{red}{\hat{\nabla}_{[a}\hat{R}_{bc]}=0}\,,\hspace{5mm} \text{where}\hspace{5mm} \hat{\nabla}_{a}=\nabla_{a}-\frac{4}{3}S_{a}\,,\hspace{5mm}\text{for}\hspace{5mm} S_{ab}{}^{c}=-\frac{2}{3}S_{[b}\delta_{a]}{}^{c}\,.
    \label{eqn:Homogeneous-Maxwell}
\end{equation}
We note once again that the above turns out to hold for a generalised action (quadratic theory~\cite{I2}), a part of which is the Einstein-Hilbert. A possible correspondence between the Faraday tensor and the Maxwell potential with the homothetic curvature and the non-metricity vector is apparent. In particular, let us focus on the correspondence $A_{a}\rightarrow Q_{a}$ and $F_{ab}\rightarrow\hat{R}_{ab}$. Taking into account that in geometrised units, $A_{a}$ and $g_{ab}$ are dimensionless, a coupling constant $k$ of length dimension is needed so that dimensional equivalence is established, i.e. 
\begin{equation}
A_{a}=kQ_{a}\hspace{5mm} \text{and} \hspace{5mm} F_{ab}=k\hat{R}_{ab}\,,
\label{eq:grav-em-equivalence}
\end{equation}
where $Q_{a}$ obviously has inverse length dimension. Thus, a potential equivalence between the homogeneous Maxwell equations and~\eqref{eqn:Homogeneous-Maxwell} is pointed out via the correspondence: $F_{ab}\rightarrow k\hat{R}_{ab}$ and $\nabla\rightarrow\hat{\nabla}$. The question is: \textit{Is there an action reproducing both Einstein and Maxwell field equations, and satisfying the condition $S_{ab}{}^{c}=-\frac{2}{3}S_{[b}\delta_{a]}{}^{c}$ (appearing in~\eqref{eqn:Homogeneous-Maxwell}) as well?} On finding such an action, the above assumed equivalence, will be established.\\

\section{Inhomogeneous Maxwell equations: From electrodynamics in curved spacetime to metric affine (gravitational) equivalent of Maxwellian electrodynamics}\label{sec:Inhomog-Maxwell-eqs}

In contrast to the homogeneous set of Maxwell equations, which springs from a purely geometric principle, the inhomogeneous one is known to be a consequence of an action principle (involving the electromagnetic field's strength and its coupling with matter.

\subsection{Maxwellian action in curved (relativistic) spacetime}\label{ssec:Maxwell-action-curved-spacetime}

Before answering the question stated in the end of the previous subsection, let us recall that the action for electrodynamics in curved (Riemannian-relativistic) spacetime, reads (e.g. see~\cite{LL1} and~\cite{DFC1})
 \begin{equation}
\mathcal{S}_{CEM}=\int{\left(R_{ab}g^{ab}+\mathcal{L}_{\text{m}}-\frac{1}{4}F_{ac}F_{bd}g^{ab}g^{cd}-A_{a}J_{b}g^{ab}\right)\sqrt{-g}\,d^{4}x}\,,
     \label{eq-Einstein-Maxwell-action}
 \end{equation}
where $J^{a}$ is the current 4-vector, $\mathcal{L}_{\text{m}}$ is the Lagrangian density of matter and $g$ the determinant of the metric tensor. In the aforementioned combined action, the electromagnetic field couples with the metric tensor of the gravitational field to form the scalar (Lorentz invariant) inner products $F_{ab}F^{ab}=F_{ac}F_{bd}g^{ab}g^{cd}$ and $A^{a}J_{a}=A_{a}J_{b}g^{ab}$. Note that in the above, there are two fundamental fields, the spacetime geometry or gravitation, and the Maxwell gauge potential. In this context, the metric tensor acts as a mediator between fields-sources of gravity-with geometric nature (vectors, tensors) and their energy content (i.e. Lagrangian densities). On the one hand, variations with respect to the potential $A_{a}$ lead to Maxwell equations of the form:
\begin{equation}
{\rm D}_{b}F^{ba}\equiv\frac{1}{\sqrt{-g}}\nabla_{b}\left(\sqrt{-g}F^{ba}\right) =J^{a}\,,\hspace{5mm}\text{where}\hspace{5mm} \frac{1}{\sqrt{-g}}\nabla_{a}\sqrt{-g}=-\frac{1}{2}Q_{a}\,.
\label{eqn:inhomog-Maxwell}
\end{equation}
The above formula reduces to $\nabla_{b}F^{ba}=J^{a}$ in Riemannian spacetime, where $Q^{a}$ vanishes.
Variations with respect to the metric field, on the other hand, lead to Einstein's equations~\eqref{eq-Einstein} and the energy-momentum tensors for the matter and Maxwell fields.

\subsection{Metric affine (gravitational) equivalent of the Maxwellian action and field equations}\label{ssec:Metric-affine-equivalent-of-Maxwell-action}

We have seen that General Relativity accommodates separate field equations for gravity and electromagnetism, which are derived by a common combined (or `coupled') action. Let us now return to our question regarding the search for an action reproducing inhomogeneous Maxwell-like equations for $\hat{R}_{ab}$, under the condition: $S_{ab}{}^{c}=-(2/3)S_{[b}\delta_{a]}{}^{c}$ (so that the homogeneous set~\eqref{eqn:Homogeneous-Maxwell} is also satisfied). Motivated by~\eqref{eq-Einstein-Maxwell-action}, the simplest action we can imagine, consists of the Einstein-Hilbert and a gravitational analogue of the Maxwellian-electromagnetic action-based on the correspondence~\eqref{eq:grav-em-equivalence}. Besides, our action (aside from the term $Q_{a}J^{a}$), is a particular case of a general model, known as \textit{quadratic theory} (e.g. see~\cite{I2},~\cite{HM} and~\cite{OVEH}). We consider the following:
\begin{equation}
    \mathcal{S}_{GEM}=\int{\left(R+\mathcal{L}_{\text{(m)}}-\frac{k^2}{4}\hat{R}_{ab}\hat{R}^{ab}-\frac{k}{2}Q_{a}J^{a}\right)\sqrt{-g}\,d^{4}x}\,,
     \label{eq-Einstein-Maxwell-action2}
\end{equation}
where $Q_{a}J^{a}$ represents a coupling between charged currents and the non-metricity vector (in analogy with the coupling $A_{a}J^{a}$ between matter and electromagnetic fields\footnote{The electric charge can be envisaged as a kind of coupling constant between matter and electromagnetic fields.}). Within the spirit of our work, unlike electromagnetic fields, we do not envisage matter (and therefore the current $J^{a}$) as a geometric quantity\footnote{Our consideration, regarding the non-geometric origin of matter, differs from the historical effort for gravito-electromagnetic unification in a metric-affine framework, started by Eddington and developed by Einstein~\cite{HG}}. Therefore, the term $Q_{a}J^{a}$ expresses a coupling between charged matter and an element of metric affine curvature. It is worth noting that no new-unknown fields are introduced, just a gravitational analogue of the classical electromagnetic action. Moreover, all action terms are invariant under general coordinate transformations (in contrast to e.g.~\cite{KPS}. Note that in the aforementioned paper the homogeneous set of Maxwell equations is not satisfied). In eq.~\eqref{eq-Einstein-Maxwell-action2} both $Q_{a}$ and therefore $\hat{R}_{ab}$ depend on the metric tensor as well as on the connection (for details see~\cite{I}). Also, we shall keep in mind that the metric appears in the Lagrangian inner products and scalars (i.e. $\hat{R}_{ab}\hat{R}^{ab}=\hat{R}_{ac}\hat{R}_{bd}~g^{ab}g^{cd}$, $Q_{a}J^{a}=Q_{a}J_{b}~g^{ab}$ and $R=R_{ab}~g^{ab}$).\\
First of all, let us consider metric variations of~\eqref{eq-Einstein-Maxwell-action2}. Taking into account the auxiliary relations in the appendix~\ref{appA11}, we arrive at Einstein field equations (of the form~\eqref{eq-Einstein}) with stress-energy tensor $T_{ab}=T^{\text{(m)}}_{ab}-(k^{2}/4)\hat{R}_{cd}\hat{R}^{cd}g_{ab}-k^{2}\hat{R}_{ac}\hat{R}^{c}{}_{b}$. Note that $R_{ab}$ and $R$ contain now contributions from torsion and non-metricity, while $T^{\text{(m)}}_{ab}$ refers to the energy-momentum tensor for matter.\\
 Regarding variations with respect to the connection (see the appendix), we receive the following field equations
 \begin{equation}
     \frac{1}{2}Q_{c}g^{ab}-Q_{c}{}^{ab}-\frac{1}{2}Q^{a}\delta^{b}{}_{c}+q^{a}\delta^{b}{}_{c}+2S_{c}g^{ab}-S^{a}\delta^{b}{}_{c}+g^{ad}S_{dc}{}^{b}+k^2\delta^{b}{}_{c}{\rm D}_{d}\hat{R}^{da}-kJ^{a}\delta^{b}{}_{c}=0\,,
     \label{eq:Gamma-FEQS}
 \end{equation}
 where ${\rm D}_{a}\equiv(1/\sqrt{-g})\nabla_{a}(\sqrt{-g}...)$. Note that the first four terms represent the so-called Palatini tensor. Moreover, all the first seven terms originate from the Einstein-Hilbert action, allowing for non-vanishing torsion and non-metricity (see chapter 2 of~\cite{I}).
 Subsequently, taking the three traces of~\eqref{eq:Gamma-FEQS}, leads to the relations:
 \begin{eqnarray}
 -\frac{3}{2}Q^{a}+3q^{a}+&4k^2{\rm D}_{b}\hat{R}^{ba}&-4kJ^{a}-4S^{a}=0\,, \hspace{10mm}
 \frac{1}{2}Q_{a}+q_{a}+k^2{\rm D}_{b}\hat{R}^{b}{}_{a}-kJ_{a}+4S_{a}=0 \nonumber \\
 &&\text{and}\hspace{5mm}\textcolor{red}{k{\rm D}_{b}\hat{R}^{ba}=J^{a}}~~(\text{with}~\textcolor{red}{{\rm D}_{a}J^{a}=0})\,.
 \label{eq:CFEQS-traces}
 \end{eqnarray}
 Note that eq.~(\ref{eq:CFEQS-traces}c) represents the inhomogeneous set of Maxwell equations. Within the same (metric-affine) framework, action~\eqref{eq-Einstein-Maxwell-action} would lead to the same equations for the Faraday field, namely $\boldsymbol{{\rm D}}_{\boldsymbol{b}}\boldsymbol{F}^{\boldsymbol{ba}}=\boldsymbol{J}^{\boldsymbol{a}}$. Let us point out that eq.~(\ref{eq:CFEQS-traces}c) is essentially a consequence of two basic mathematical properties and one physical property. In detail, the two mathematical properties are: firstly, the similar mathematical construction between the homothetic curvature $\hat{R}_{ab}$ and the Faraday $F_{ab}$ tensor field (i.e. written as the gradient of a vector field); secondly, the linear dependence of the non-metricity vector $Q_{a}$ on the connection, so that $\delta_{\Gamma}Q_{a}=2\delta_{a}{}^{d}\delta_{c}{}^{b}\delta\Gamma^{c}{}_{bd}$. The physical property is associated with the action~\eqref{eq-Einstein-Maxwell-action2} itself.

\subsection{Constraints and similarities with classical unified theories}\label{ssec:Constraints}
 
 We observe that charge conservation is expressed in the form ${\rm D}_{a}J^{a}=0~(\Leftrightarrow \nabla_{a}J^{a}=(1/2)Q_{a}J^{a})$. Moreover, taking the nabla divergence of~(\ref{eq:CFEQS-traces}c), we find out the constraint
 \begin{equation}
 \nabla_{a}J^{a}=\frac{k}{2}\left(\hat{R}_{ab}\hat{R}^{ab}+Q_{a}\nabla_{b}\hat{R}^{ba}\right)\hspace{5mm} \text{or}\hspace{5mm} k\hat{R}_{ab}\hat{R}^{ab}=Q_{a}\left(J^{a}-k\nabla_{b}\hat{R}^{ba}\right)\,.
 \label{eqn:constraint}
\end{equation}
 In other words, we have figured out that the last two terms in the action~\eqref{eq-Einstein-Maxwell-action2} are actually related with each other through the above expression. \\
 Subsequently, considering various combinations of the three traces in~\eqref{eq:CFEQS-traces} with the initial field equations~\eqref{eq:Gamma-FEQS} (this involves some lengthy but straightforward algebra)\footnote{Note that due to the non-metricity requirement, raising indices is no-longer a trivial operation. For instance, raising indices in~(\ref{eq:CFEQS-traces}b) leads to
 \begin{equation}
     \frac{1}{2}Q^{a}+q^{a}-kQ_{b}{}^{ca}\hat{R}^{b}{}_{c}+\frac{k^{2}}{2}Q_{b}\hat{R}^{ba}+4S^{a}=0\,.\nonumber
 \end{equation}
 }, we eventually arrive at the constraints
 \begin{equation}
     Q^{a}=4q^{a}=-\frac{16}{3}S^{a}\,.
     \label{eqn:constraints2}
 \end{equation}
 Namely, within the framework of the action~\eqref{eq-Einstein-Maxwell-action2}, the non-metricity and torsion vectors are linearly dependent, so that they all together correspond to only one degree of freedom. The same thing generally happens when considering only the Einstein-Hilbert action (e.g. see~\cite{I}). In particular, it is well-known that Einstein-Hilbert action does not reproduce general relativity. Instead, it leads to Einstein's field equations along with an additional degree of freedom expressed by~\eqref{eqn:constraints2}. As a consequence of the latter, relation~\eqref{eq:grav-em-equivalence} recasts into
 \begin{equation}
     A_{a}=kQ_{a}=4kq_{a}-\frac{16}{3}kS_{a}\hspace{5mm}\text{and}\hspace{5mm} F_{ab}=k\hat{R}_{ab}=4kq_{ab}=-\frac{16}{3}kS_{ab}\,,
     \label{eq:gravito-electrom-equiv2}
 \end{equation}
 where $q_{ab}\equiv \partial_{[a}q_{b]}$ and $S_{ab}\equiv \partial_{[a}S_{b]}$. In other words, the vectorial degree of freedom expressed by~\eqref{eqn:constraints2} and allowed by the Einstein-Hilbert action provides a gravitational equivalent for the Maxwell field. Furthermore, following some lengthy operations, involving eqs.~\eqref{eq:Gamma-FEQS} and~\eqref{eq:CFEQS-traces} (see Ch. 2 of~\cite{I}), it can be shown that the torsion and non-metricity tensors are related with the associated vectors via
 \begin{equation}
     S_{ab}{}^{c}=-\frac{2}{3}S_{[b}\delta_{a]}{}^{c}\hspace{5mm}\text{and}\hspace{5mm} Q_{abc}=\frac{1}{4}Q_{a}g_{bc}\,.
     \label{eqn:tensor-vector-S-Q-constraints}
 \end{equation}
 The above constraints hold exactly the same for action~\eqref{eq-Einstein-Maxwell-action2}, given that eq~\eqref{eq:Gamma-FEQS} reduces to Einstein-Hilbert $\Gamma$-field equations under~(\ref{eq:CFEQS-traces}c). Therefore, the homogeneous set of Maxwell equations in the form of~\eqref{eqn:Homogeneous-Maxwell}, is also satisfied by the $\hat{R}_{ab}$ field in the case of the action we examine.\\

\section*{Closing remarks-Questions for further research}\label{sec:Discussion}

Although the present work was initially motivated by the problem of classical gravito-electromagnetic unification, our study points out more a potential equivalence between the Maxwell field and a metric affine component of the gravitational field (i.e. homothetic or length curvature), analogous to mass-energy equivalence. Within the unified theories context, the present work moves between the lines of Weyl and Eddington-Einstein. It shares some similarities with both the aforementioned approaches but it essentially differs from both. In fact, envisaging electromagnetism as a component of metric-affine gravity, dates back to the efforts of Weyl, Eddington and Einstein~\cite{HG},~\cite{LOR} (refer to the aforementioned reviews for any information concerning past efforts and failures of unification).

 Overall, we have shown that the antisymmetric part of the Ricci tensor, namely the homothetic curvature, satisfies all of Maxwell equations. This finding points out the fundamental question: \textit{Is it possible to exist two different kinds of fields both satisfying Maxwell equations and describing different things? If not, should electromagnetism be envisaged as a form, instead of a source, of gravity on large scales? Alternatively, are electromagnetic fields equivalent to gravitational fields and which is the equivalence relation?} Our work shows that there must be such an equivalence, taking the form of~\eqref{eq:grav-em-equivalence}, so that the Maxwell field can be calculated from a given metric. The aforementioned relation implies that a given electromagnetic field has a gravitational equivalent determined via the conversion constant $k$. It is worth noting that there is a remarkable analogy between gravito-electromagnetic (eq.~\eqref{eq:grav-em-equivalence}) and mass-energy equivalence, i.e. $E=mc^2$ ($k$ is the counterpart of $c^{2}$). Showing the existence of a \textit{gravito-electromagnetic equivalence} is essentially 
 the contribution of the present piece of work. Therefore, two crucial questions arise.
 
  Firstly, which is the nature of the conversion constant $k$ and how can it be determined? Let us make a \textit{conjecture}. On the one hand, we observe that the action term $J_{a}Q^{a}$, introduced in~\eqref{eq-Einstein-Maxwell-action2}, establishes a coupling between matter and non-metricity, mediated by the electric charge. On the other hand, within classical electrodynamics, the electric charge is known to act as a coupling constant between matter and electromagnetic field (see $J_{a}A^{a}$ in~\eqref{eq-Einstein-Maxwell-action}). The aforementioned double coupling potentially implies an equivalence relation between non-metricity and the Maxwell field, where the electric charge plays the role of the coupling constant. Besides, we take into account that the electric charge has length dimension in geometrised units. Therefore and in other words, we state the following question: Could the coupling constant $k$ (with length geometrised dimension) be identified as the total electric charge of a given charged distribution? If this is the case, it would appear that the electric charge behaves on large-scales as a quantity which translates a given electromagnetic field into its gravitational equivalent. Furthermore, according to~\eqref{eq:grav-em-equivalence}, with $k\rightarrow\mathcal{Q}$, opposite charges correspond to homothetic curvature of opposite sign. Could the macroscopic interaction between a positive and a negative charge distribution be envisaged as a consequence of an `interaction' between opposite kinds of homothetic curvature?
 
 Secondly, how are the properties of the Maxwell field ($A_{a}=kQ_{a}=(16/3)kS_{a}$, via~\eqref{eqn:constraints2}) reconciled with the geometric significance/properties of non-metricity and torsion? The aforementioned properties are respectively the change to a vector's magnitude under its parallel transport along a given curve, and the impossibility to form a closed (small) parallelogram under parallel transport of one vector to the direction of another~\cite{I}.\\
 Addressing the above exposed questions/problems will be an object of my future efforts.\\ 

\begin{appendix}

\section{Metric and connection variations}\label{appA11}

In deriving the field equations within the main text we make use of the following relations for metric and connection variations~\cite{I},~\cite{I2}. Concerning the former, we have:
\begin{equation}
    \delta_{g}Q_{a}=\partial_{a}\left(g_{bc}\delta g^{bc}\right)\,,\hspace{4mm} \delta_{g}\hat{R}_{ab}=\partial_{[a}\delta_{g}Q_{b]}=0\,,\hspace{4mm}\delta_{g}\sqrt{-g}=-(1/2)\sqrt{-g}g_{ab}\delta g^{ab}\,,\nonumber
\end{equation}
\begin{equation}
\delta_{g}\left(\hat{R}_{ab}\hat{R}^{ab}\right)=\delta_{g}\left(\hat{R}_{ab}\hat{R}_{cd}g^{ac}g^{bd}\right)=-2\hat{R}_{ac}\hat{R}^{c}{}_{b}\delta g^{ab}\nonumber
\end{equation}
and
\begin{equation}
    \delta_{g}\left(\hat{R}_{ab}\hat{R}^{ab}\sqrt{-g}\right)=\left(-2\hat{R}_{ac}\hat{R}^{c}{}_{b}-(1/2)\hat{R}_{cd}\hat{R}^{cd}g_{ab}\right)\sqrt{-g}\\\delta g^{ab}\,.\nonumber
\end{equation}
As for the latter, we deploy:
\begin{equation}
    \delta_{\Gamma}Q_{a}=2\delta_{a}{}^{d}\delta_{c}{}^{b}\delta\Gamma^{c}{}_{bd}\hspace{4mm}\text{and}\hspace{4mm}\delta_{\Gamma}(\hat{R}_{ab}\hat{R}^{ab})=-4\nabla_{b}\hat{R}^{ba}\delta\Gamma^{c}{}_{ca}=-4\nabla_{d}\hat{R}^{da}\delta^{b}{}_{c}\delta\Gamma^{c}{}_{ab}\,,\nonumber
\end{equation}
with $\delta_{a}{}^{b}$ being the Kronecker symbol.\\\\

\end{appendix}

\textbf{Acknowledgements:} To my beloved friend Evangelos Kipouridis, with whom I have been sharing my thoughts and interest in physics since I was at school.\\
I am grateful to Damos Iosifidis for providing me with many useful technical details regarding the metric affine formulation; for some illuminating discussions and his overall great willingness to help. I also thank prof Tomi S. Koivisto for his insightful comments.\\
The present research work was supported by the Hellenic Foundation for Research and Innovation (H.F.R.I.), under the ‘Third Call for H.F.R.I. PhD Fellowships’ (Fellowship No. 74191).\\\\

\end{document}